\documentclass[12pt,onecolumn]{article}
\usepackage[dvips]{graphicx}
\usepackage{amsfonts}
\usepackage{indentfirst}

\title{Information-Theoretic Comparison of Quantum Many-Body Systems}

\author{ ~K.~Ch.~Chatzisavvas\footnote{\texttt{e-mail:\,
         kchatz\,@\,auth.gr}},
         ~C.~P.~Panos\footnote{\texttt{e-mail:\,
         chpanos\,@\,auth.gr}},
         ~S.~E.~Massen\footnote{\texttt{e-mail:\,
         massen\,@\,physics.auth.gr}}
\\
 {\it  Physics Department,}\\
        {\it Aristotle University of Thessaloniki,}\\
                {\it  54124 Thessaloniki, Greece}
 }

\date{May, 2003}

\begin{document}

\maketitle

\begin{abstract}
An information measure inspired by Onicescu's information energy
and Uffink's information measure (recently discussed by Brukner
and Zeilinger) are calculated as functions of the number of
particles $N$ for fermionic systems (nuclei and atomic clusters)
and correlated bosonic systems (atoms in a trap). Our results are
compared with previous ones obtained for Shannon's information
entropy, where a universal property was derived for atoms, nuclei,
atomic clusters and correlated bosons. It is indicated that
Onicescu's and Uffink's definitions are finer measures of
information entropy than Shannon's.
\end{abstract}

\vspace{1.0cm}

Onicescu \cite{Onicescu} introduced the concept of information
energy $E$ as a finer measure of dispersion distributions than
that of Shannon's information entropy \cite{Shannon1, Shannon2}.
So far, only the mathematical aspects of this concept have been
developed, while the physical aspects have been neglected
\cite{Agop}.

The information energy for a single statistical variable $x$ with
the normalized density $\rho(x)$ is defined by
\begin{equation}\label{eq:eq1}
  E(\rho)=\int \rho^2(x)\,dx
\end{equation}
For a Gaussian distribution of mean value $\mu$, standard
deviation $\sigma$ and normalized density
\begin{equation}\label{eq:eq2}
  \rho(x)=\frac{1}{\sqrt{2\pi}\sigma}\,e^{-\frac{(x-\mu)^2}{2\sigma^2}}
\end{equation}
relation (\ref{eq:eq1}) gives
\[
  E=\frac{1}{2\pi\sigma^2}\int_{-\infty}^{\infty}e^{-\frac{(x-\mu)^2}{\sigma^2}}\,dx
\]
Thus
\begin{equation}\label{eq:eq3}
  E=\frac{1}{2\sigma\sqrt{\pi}}
\end{equation}

Therefore, the greater the information energy $E$, the narrower
the Gaussian distribution. $E$ does not have the dimension of
energy, but it has been connected with Planck's constant appearing
in Heisenberg's uncertainty relation \cite{Agop, Ioannidou}.

For a 3-dimensional spherically symmetric density distribution
$\rho(r)$ the obvious generalization of (\ref{eq:eq1}) is
\begin{equation}\label{eq:eq4}
  E_r=\int\rho^2(r)4\pi r^2\,dr
\end{equation}
and
\begin{equation}\label{eq:eq5}
  E_k=\int n^2(k)4\pi k^2\,dk
\end{equation}
in position- and momentum-space respectively, where $n(k)$ is the
density distribution in momentum-space.

$E_r$ has the dimension of inverse volume, while $E_k$ of volume.
Thus the product $E_r E_k$ is dimensionless and is a measure of
the concentration (or the information content) of the density
distribution of a quantum system. As seen from (\ref{eq:eq3}) $E$
increases as $\sigma$ decreases (or the concentration increases)
and Shannon's information entropy (or uncertainty) $S$ decreases.
Clearly, Shannon's information $S$ and information energy $E$ are
reciprocal. In order to be able to compare them, we define the
quantity
\begin{equation}\label{eq:eq6}
  S_E=\frac{1}{E_r E_k}
\end{equation}
as a measure of the information content of a quantum system in
both position and momentum spaces.

In place of Shannon information, Brukner and Zeilinger
\cite{Brukner} propose the quantity
\begin{equation}\label{eq:eq7}
  I=\mathcal{N}\sum_{i=1}^{n}(p_i-\frac{1}{n})^2
\end{equation}
from which they derive their notion of information content of a
discrete probability distribution $p_1, p_2, \ldots, p_n$. The
quantity $\sum_{i=1}^{n}(p_i-\frac{1}{n})^2$ is one of the class
of measures of the concentration of a probability distribution
given by Uffink \cite{Uffink, Maassen}. For a continous
3-dimensional density distribution $\rho(r)$, relation
(\ref{eq:eq7}) is extended as ($\mathcal{N}=1$)
\begin{equation}\label{eq:eq8}
  I_r=\int\Big( \rho(r)-\tilde{\rho}(r)\Big)^2 4\pi r^2\,dr
\end{equation}
and
\begin{equation}\label{eq:eq9}
  I_k=\int\Big( n(k)-\tilde{n}(k)\Big)^2 4\pi k^2\,dk
\end{equation}
in position- and momentum space respectively, $\tilde{\rho}(r)$ is
the equivalent uniform distribution defined according to the
relation
\begin{equation}\label{eq:eq10}
 \tilde{\rho}(r)=\left\{ \begin{array}{ll}
  \rho_0 & 0<r<R_U \\
  0 & r>R_U
  \end{array} \right.
\end{equation}
where $\rho_0$=constant and $R_U$=$R_{\rm{uniform}}$ are fixed by
the relation

\begin{equation}\label{eq:eq11}
  {\langle r^2 \rangle}_U={\langle r^2 \rangle}_{\rho(r)}
\end{equation}
where
\begin{equation}\label{eq:eq12}
  {\langle r^2 \rangle}_U=\int_0^{R_U}\rho_0 r^2 4\pi r^2\,dr
\end{equation}
and
\begin{equation}\label{eq:eq13}
  {\langle r^2 \rangle}_{\rho(r)}=\int_0^{\infty}\rho(r) r^2 4\pi r^2\,dr
\end{equation}
while
\begin{equation}\label{eq:eq14}
  R_U=\sqrt{\frac{5}{3}<r^2>_{\rho(r)}}
\end{equation}
and
\begin{equation}\label{eq:eq15}
  \rho_0=\frac{3}{4\pi R_U^3}
\end{equation}
$\tilde{n}(k)$ is the equivalent uniform distribution in
momentum-space, defined in a similar way. Thus we define a measure
of information content by the relation
\begin{equation}\label{eq:eq16}
  S_I=\frac{1}{I_r I_k}
\end{equation}
which gives (\ref{eq:eq6}) putting
$\tilde{\rho}(r)=\tilde{n}(k)=0$.

We calculate $S_E$ and $S_I$ as functions of the number of
particles $N$ for three quantum many-body systems, where $\rho(r)$
and $\eta(k)$ are calculated numerically:
\begin{enumerate}
\item Nuclei, using the Skyrme III parametrization of the nuclear
field \cite{Dover}. Here $N$ is the number of nucleons in nuclei.
\item Atomic clusters, employing a Woods-Saxon potential
parametrized by Ekardt \cite{Ekardt}. Here $N$ is the number of
valence electrons. \item A correlated bosonic system (atoms in a
trap) \cite{Moustakidis, Fabrocini}. Here $N$ is the number of
atoms in the trap.
\end{enumerate}

In Fig.1 we plot $S_E$ as a function of ${N}$ for nuclei and
clusters and in Fig.2 $S_I(N)$ for the same systems. In Fig.3 we
plot $S_E(N)$ and in Fig.4 $S_I(N)$ for a correlated bosonic
system. It is seen that $S_E$ depends linearly on $N$ for both
nuclei and atomic clusters. Also $S_I$ shows a similar trend (a
power of $N$) for nuclei and clusters. However the dependence
$S_E(N)$ and $S_I(N)$ is different for correlated bosons compared
with nuclei and clusters.

Our fitted expressions are:
\[
  S_E(\textrm{clusters})=143.420\,N, \quad
  S_E(\textrm{nuclei})=73.883\,N
  \quad \,\textrm{(Fig.1)}
\]

\[
  S_I(\textrm{clusters})=431.576\,N^{1.719}, \quad
  S_I(\textrm{nuclei})=260.275\,N^{1.554}
  \quad \,\textrm{(Fig.2)}
\]
We can compare with the universal relation $S(N)=a+b\ln{N}$ ($a,
b$ are constants depending on the system) obtained recently
\cite{Massen2} for Shannon's information entropy for fermionic
systems (atoms, nuclei and atomic clusters) and correlated bosonic
systems \cite{Moustakidis} (atoms in a trap). It was seen
\cite{Massen2} that $S(N)$ shows the same dependence on $N$ for
all the systems considered i.e. nuclei, clusters, atoms and
correlated bosons.

It is conjectured that nuclei and atomic clusters are equivalent
from an information-theoretic point of view in the following
sense: under any definition of information content (e.g. Shannon,
Onicescu or Uffink), the dependence of information shows a similar
trend (linear on $\ln{N}$ for Shannon, linear on $N$ for Onicescu
and a power of $N$ for Uffink). However, the similarity breaks
down for bosons. This indicates that $S_E$ and $S_I$ distinguish
between fermions and correlated bosons i.e. they are  finer
measures of infomation than Shannon's $S$. Our results may
contribute to the recent debate between Brukner-Zeilinger and
Timpson for a possible inadequacy of the Shannon information
\cite{Brukner, Timpson}

\newpage

\begin{figure}[ht]
\centering
\includegraphics[height=7.0in]{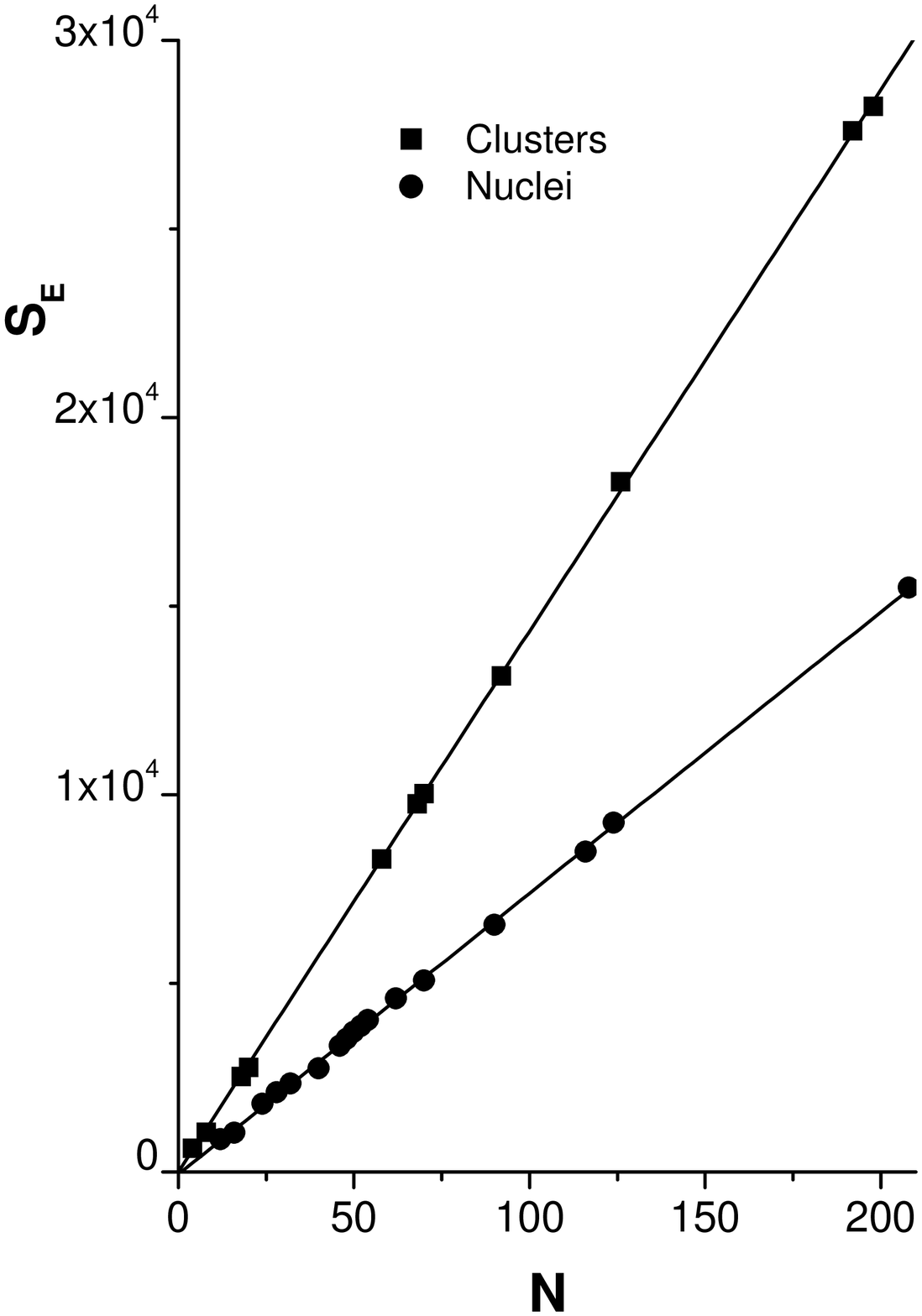}
\caption{Onisescu's information entropy $S_E$ as function of $N$
\& for nuclei (circles) and atomic clusters (squares)}
\end{figure}

\newpage

\begin{figure}[ht]
\centering
\includegraphics[height=7.0in]{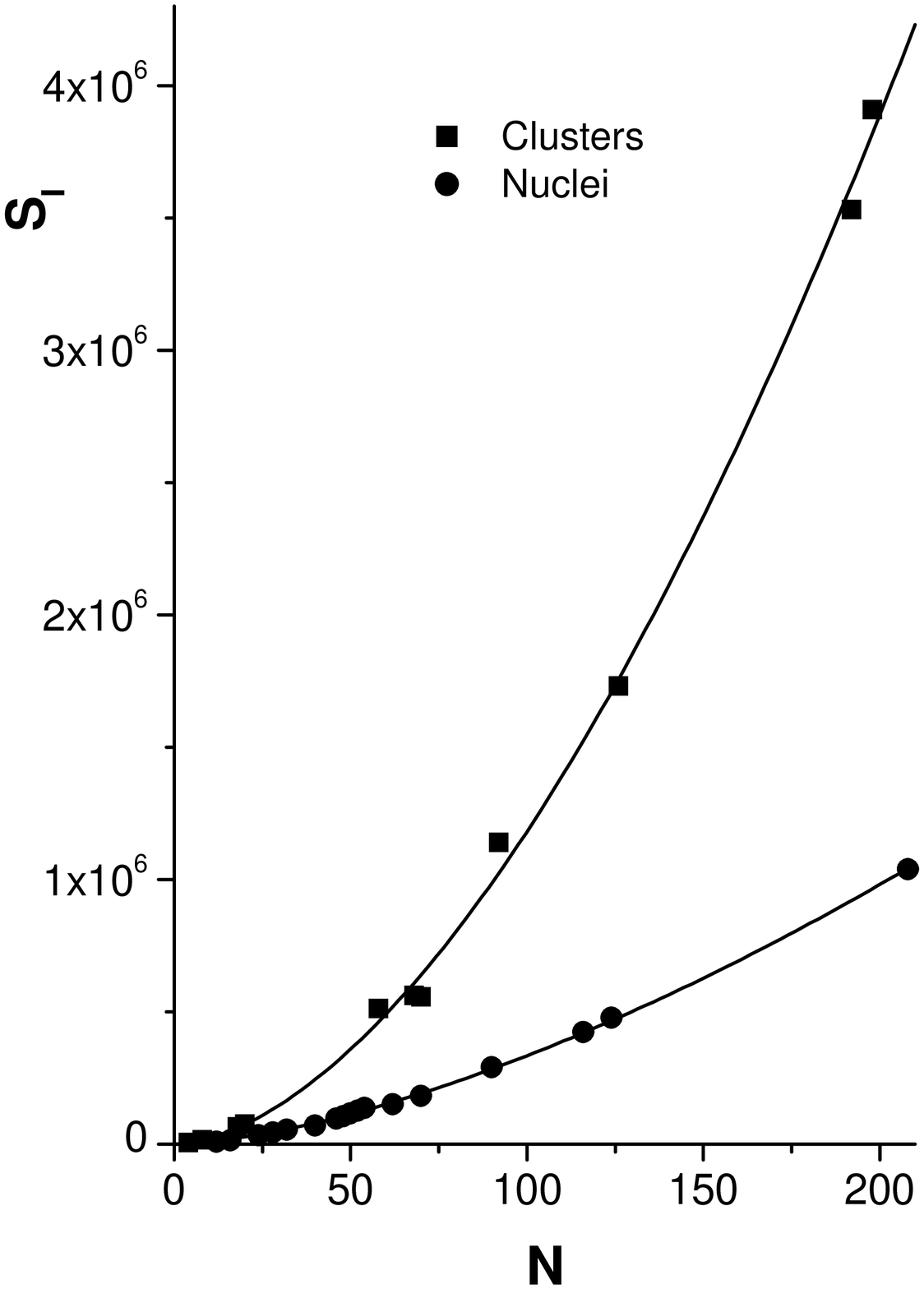}
\caption{The same as in Fig.1 but for Uffink's information
      \& entropy $S_I$}
\end{figure}

\newpage

\begin{figure}[ht]
\centering
\includegraphics[height=7.0in]{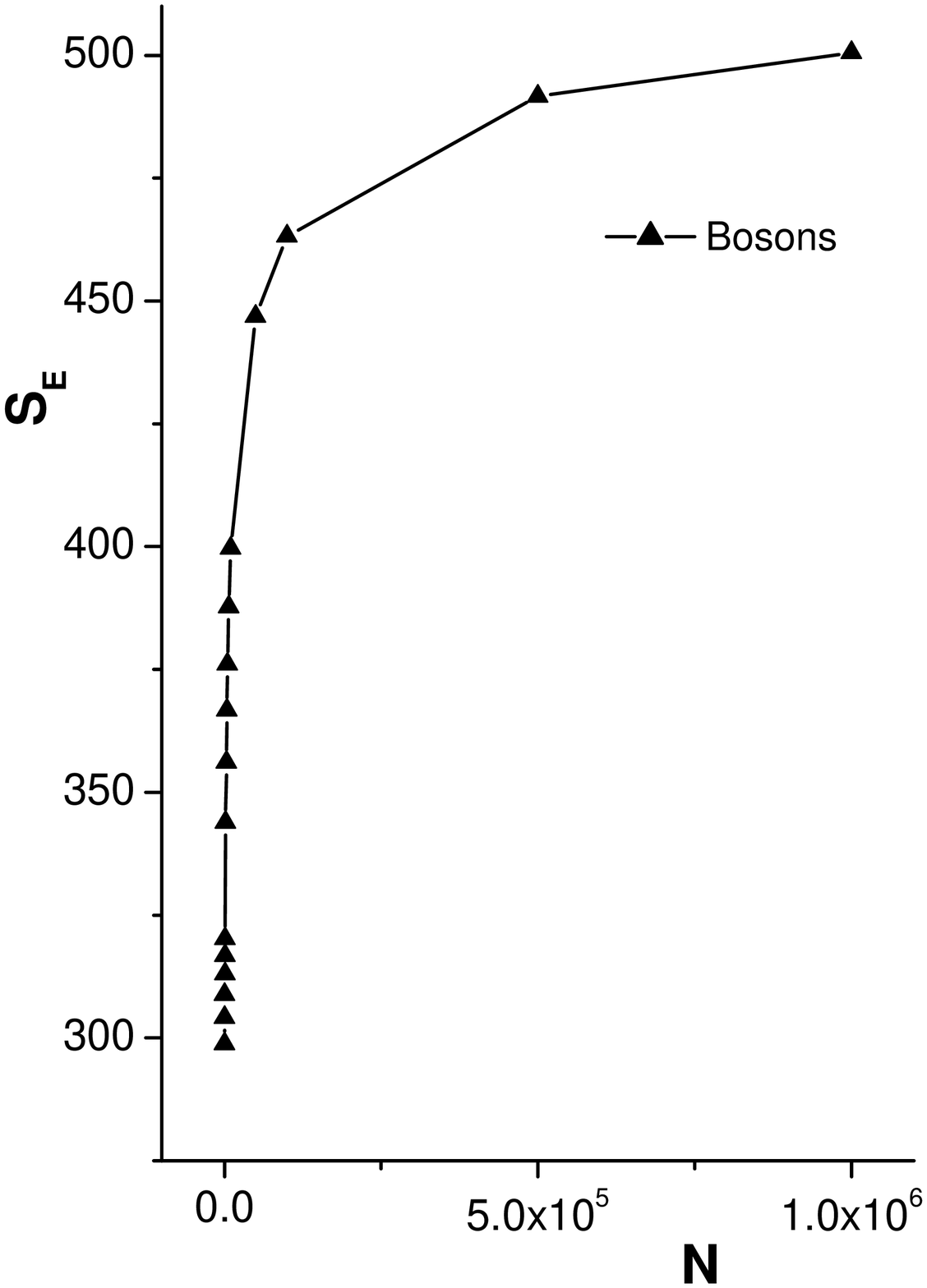}
\caption{Dependence of $S_E$ on $N$ for correlated bosons}
\end{figure}

\newpage

\begin{figure}[ht]
\centering
\includegraphics[height=7.0in]{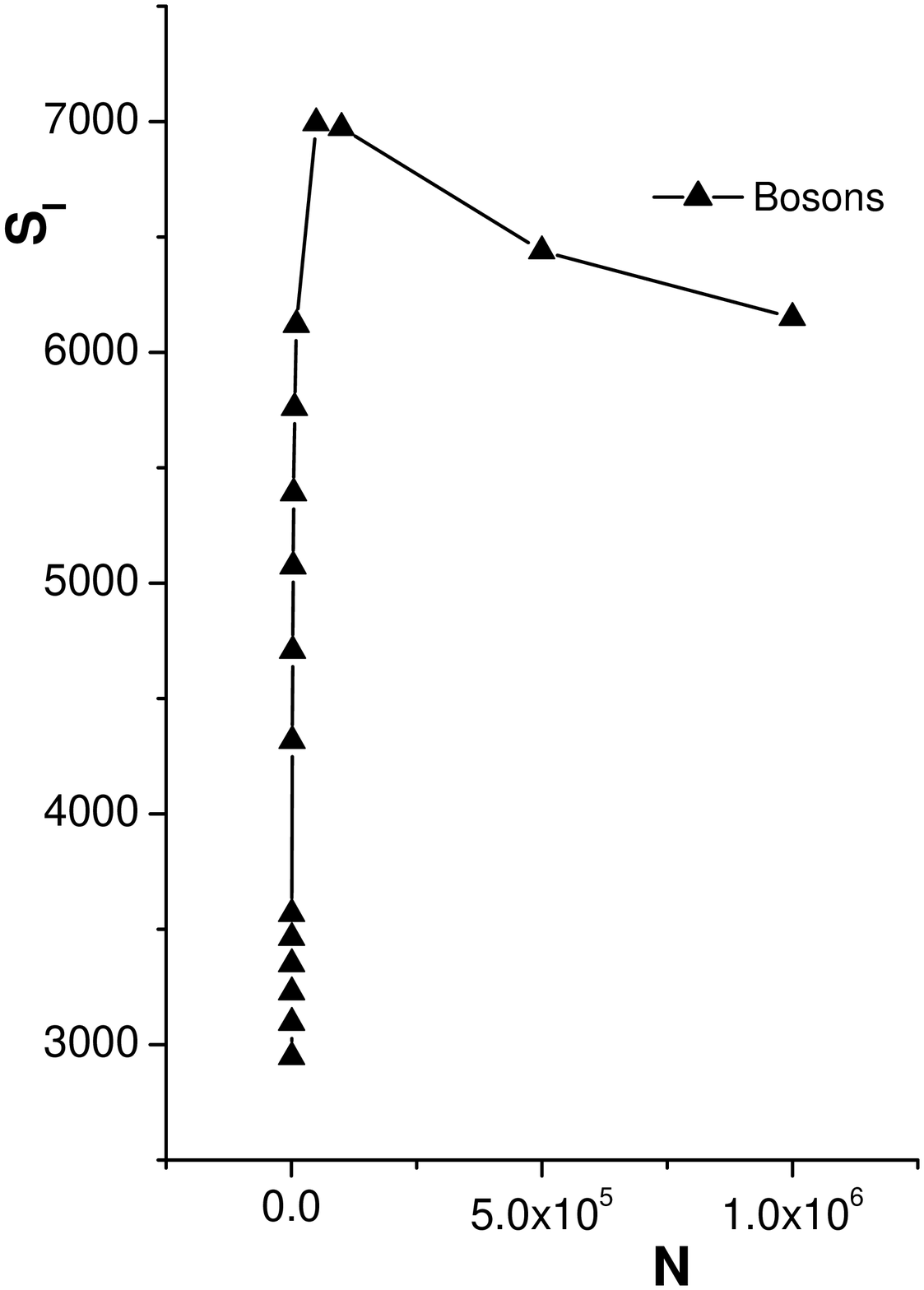}
\caption{The same as in Fig.3 but for $S_I$}
\end{figure}


\begin{thebibliography}{99}

\bibitem{Onicescu}
 O.~Onicescu, \emph{C. R. Acad. Sci. Paris A} \textbf{263} (1966)

\bibitem{Shannon1}
 C.~E.~Shannon, \emph{Bell Syst. Tech.} \textbf{27} (1948), 379;
 ibid. \textbf{27} (1948) 623

\bibitem{Shannon2}
 C.~E.~Shannon, W.~Weaver, \emph{The Mathematical Theory of
 Communication}, University of Illinois Press, Urbana and Chicago,
 1963

\bibitem{Agop}
 M.~Agop, C.~Buzea, C.~Gh.~Buzea, L.~Chirila, S.~Oancea,
 \emph{Chaos, Solitons and Fractals} \textbf{7} (1996), 659

\bibitem{Ioannidou}
 H.~Ioannidou, \emph{Int. Journ. of Theor. Phys. 20} \textbf{20} (1981),
 1129

\bibitem{Brukner}
 C.~Brukner, A.~Zeilinger, \emph{Phys. Rev.A} \textbf{63} (2001),
 022113

\bibitem{Uffink}
 J.~Uffink, \emph{PhD Thesis}, University of Utrecht (1990)

\bibitem{Maassen}
 H.~Maassen, J.~Uffink, \emph{Phys. Rev. Lett.} \textbf{60}
 (1988), 1103

\bibitem{Dover}
C.~B.~Dover, N.~Van ~Giai, \emph{Nucl. Phys. A} \textbf{190}
(1972), 373

\bibitem{Ekardt}
N.~Ekardt, \emph{Phys. Rev. B} \textbf{29} (1984), 1558

\bibitem{Moustakidis}
S.~E.~Massen, Ch.~C.~Moustakidis, C.~P.~Panos, \emph{Phys. Lett.
A} \textbf{299} (2002), 131

\bibitem{Fabrocini}
A.~Fabrocini, A.~Polls, \emph{Phys. Rev. A} \textbf{60} (1999),
2319

\bibitem{Massen2}
S.~E.~Massen, C.~P.~Panos, \emph{Phys. Lett. A} \textbf{246}
(1998), 530

\bibitem{Timpson}
C.~G.~Timpson, LANL archive, quant-ph/0112178

\end{thebibliography}
\end{document}